\documentclass[aps,pre,twocolumn,amsmath,amssymb,superscriptaddress,showpacs,floatfix,nofootinbib]{revtex4-2} 

\usepackage{graphicx}
\usepackage{subfigure}
\usepackage{lipsum}
\usepackage{hyperref}
\usepackage[capitalise]{cleveref}
\usepackage{amsmath}
\usepackage[usenames,dvipsnames]{xcolor}
\hypersetup{
    colorlinks,
    linkcolor={blue!50!black},
    citecolor={blue!50!black},
    urlcolor={blue!80!black}
}
\usepackage{array}

\DeclareMathOperator*{\argmax}{argmax}

\newcommand{\taustick}{\tau_0}
\newcommand{\tauunstick}{\tau_1}

\begin{document}
\title{Hidden Markov modeling of single particle diffusion with stochastic tethering}

\author{Amit Federbush}
\affiliation{Department of Condensed Matter Physics, Tel Aviv University}
\affiliation{The Center for Physics and Chemistry of Living Systems, Tel Aviv University}
\author{Amit Moscovich}
\affiliation{Department of Statistics and Operations Research, Tel Aviv University}
\author{Yohai Bar-Sinai}
\affiliation{Department of Condensed Matter Physics, Tel Aviv University}
\affiliation{The Center for Physics and Chemistry of Living Systems, Tel Aviv University}

\begin{abstract}
The statistics of the diffusive motion of particles often serve as an experimental proxy for their interaction with the environment. However, inferring the physical properties from the observed trajectories is challenging. Inspired by a recent experiment, here we analyze the problem of particles undergoing two-dimensional Brownian motion with transient tethering to the surface. We model the problem as a Hidden Markov Model where the physical position is observed and the tethering state is hidden. We develop an alternating maximization algorithm to infer the hidden state of the particle and estimate the physical parameters of the system. The crux of our method is a saddle-point-like approximation, which involves finding the most likely sequence of hidden states and estimating the physical parameters from it. Extensive numerical tests demonstrate that our algorithm reliably finds the model parameters, and is insensitive to the initial guess. We discuss the different regimes of physical parameters and the algorithm's performance in these regimes. We also provide a free software implementation of our algorithm.
\end{abstract}

\maketitle

\section{\label{sec:introduction}Introduction}
Since the early days of statistical mechanics, the statistics of the stochastic motion of mesoscopic particles are an important experimental probe for their microscopic properties. Most prominently, the Gaussian statistics of Brownian motion provided experimental proof of the atomic nature of matter~\cite{Einstein1905,Smoluchowski1906, Perrin1909} and was used to measure Avogadro's number~\cite{kappler1931}. To this day, new models are actively developed to explain deviations from purely Brownian statistics in biological and colloidal systems~\cite{Wang2009, Wang2012}. Of specific interest are systems undergoing anomalous diffusion which exhibit a Fickian behavior, i.e.~a mean-squared displacement that is linear with time where the displacement statistics are non-Gaussian~\cite{Wang2009,Wang2012,Chechkin2017}. These deviations from Gaussianity can serve as an accessible probe for various experiments \cite{Guan2014, Chakraborty2019, Chakraborty2020, Pastore2022, Rusciano2022, Ciarlo2023}.

\begin{figure}
\centering
\includegraphics[width=\linewidth]{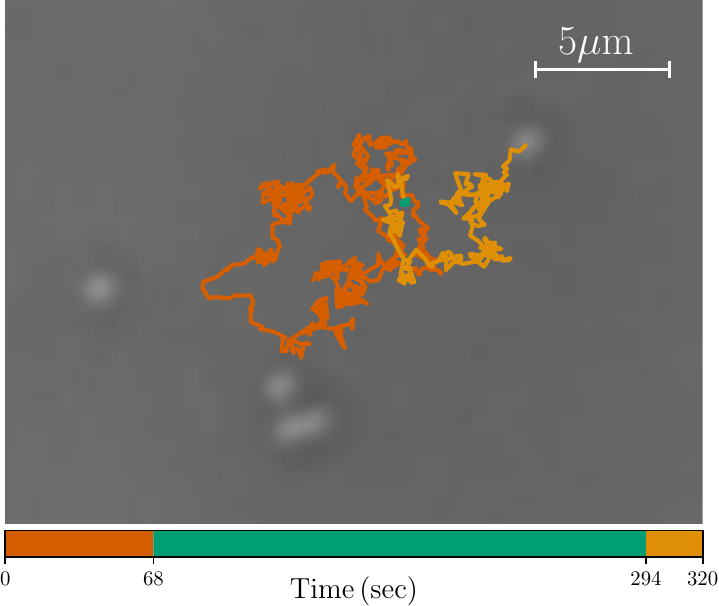}
\caption{A snapshot of an experimental microscopy video of peptide-coated microparticles, in a system similar to that of Chakraborty et. al. \cite{Chakraborty2019}. The colored curve represents the top-right particle's recorded trajectory in the 320 seconds preceding the snapshot. The trajectory is colored based on time, as indicated by the timeline at the bottom. During the time segment $t \in [68,294]$ the particle is confined to a small region of space (shown in green) due to tethering. Image courtesy of Amandeep Sekhon, Roy Beck, and Yael Roichman.}
\label{fig: experimental trajectory}
\end{figure}

In the analysis of such systems, two key sources for non-Gaussianity are considered. First, particles may undergo several different types of diffusion modes, stochastically switching between them~\cite{YoshinaIshii2006, Das2009, Ott2013, Thapa2018, Granik2019, Falcao2020}. Second, particles may be transiently confined to a small region~\cite{Xu2011,Skaug2013,Bernstein2016,Wang2018,Slator2018,Callegari2019,Chakraborty2019,Kowalek2019,Jensen2019,Huseyin2021,Doerries2022_PRE,Doerries2022_Interface,Simon2023,Doerries2023}. In this work, inspired by the experimental system of Chakraborty et.~al.~\cite{Chakraborty2019}, we focus on the latter, and specifically on two-dimensional (2D) diffusion with transient tethering to the underlying surface. \cref{fig: experimental trajectory} depicts an example trajectory from such an experiment.
In such experiments, colloids or nano-particles are coated with molecules of interest (typically, peptides) and undergo 2D diffusion on a surface coated with a different molecule. The interaction between the molecules leads to stochastic transient tethering of the particles to the surface. The experiment aims to extract information regarding the interaction between the peptides from the frequency of these tethering and untethering events.

However, identifying these events may be challenging, since the tethering to the surface is not directly observed, but rather needs to be inferred from the observed trajectories. Several computational methods have been suggested to tackle this in the past~\cite{Bernstein2016, Slator2018, Simon2023}. In this work, we present a simple and computationally efficient algorithm to infer tethering and untethering events from observed trajectories, to estimate the tethering and untethering rates, the diffusion coefficient, and the effective confinement area of the interaction potential. The algorithm is developed under the assumption that the dynamics follow normal diffusion with Poissonian tethering/untethering events, but can be easily extended to other types of diffusion and tethering/untethering rates.

The structure of this manuscript is as follows: In \cref{sec:model} we describe the stochastic hidden Markov model which we use to model the problem, discuss the different time scales that the model introduces, and define the regimes of model parameters where our methods are valid. In \cref{sec:problem algorithm} we describe our algorithm and its underlying approximations, and in \cref{sec:results} we present our numerical results on synthetic data. Finally, in \cref{sec:discussion} we discuss the results and possible generalizations. The full source code of the implemented algorithm is available on GitHub~\cite{github}.

\section{Stochastic Model}
\label{sec:model}
Our Markov model describes a particle alternating between a freely diffusing state and a tethered state in 2D. The transition between the states is modeled as a standard two-state continuous-time Markov chain with characteristic times $\taustick$ and $\tauunstick$, which are the inverses of the average rates of tethering and untethering events, respectively. 
We denote the state of the particle by $S(t)$ where $S=0$ corresponds to the free state and $S=1$ to the tethered state.
In the free state, the particle undergoes standard Brownian motion, while in the tethered state, it is also confined to a harmonic potential centered at the tether point $X^*(t)$. Explicitly, the position of the particle $X(t)$ follows an Ornstein-Uhlenbeck process with a state-dependent potential:
\begin{align}
        \dot X(t) &= -\frac{k}{\gamma}(X(t)-X^*(t)) S(t) + \sqrt{2D}\xi(t),
\end{align}
where $k$ is the spring constant, $\gamma$ is the friction coefficient, $S(t) \in \{0, 1\}$ indicates the free/tethered state, $D$ is the diffusion coefficient, and $\xi$ is a standard two-dimensional white noise $\langle \xi_i(t)\xi_j(t')\rangle=\delta_{ij}\delta(t-t')$. Observe that only the ratio $k/\gamma$ plays a role in the model, rather than $k$ and $\gamma$ individually.
We assume the particle is tethered to the point at which the transition $S=0\to 1$ occurs. That is, if the particle becomes tethered at time $t_1$, and remains tethered until untethering at $t_2$,  then $X^*(t)$ satisfies
\begin{align} \label{eq:continuous xstar}
    X^*(t)=X(t_1) \qquad \forall t \in [t_1,\ t_2).
\end{align} 
To summarize, the particle dynamics are modeled by a continuous-time Markov process,
\begin{align}
    F(t):= \big(X(t),\ S(t),\ X^*(t)\big).
\end{align}
The model is specified by 4 parameters: $\taustick, \tauunstick, D$ and the ratio $k/\gamma$. It is more convenient to work with the variable $A=(D\gamma)/k$ which is the characteristic  area that the particle explores in the tethered state.
For notational purposes, we group the model parameters as $\Theta := (\taustick,\tauunstick,D,A)$. Lastly, we note that $\frac{A}{D}=\frac{\gamma}{k}$ has units of time. Its meaning is discussed below. 

The transition probabilities of $S$ are Poissonian.
In the untethered state $S(t) = 0$, the particle undergoes classic Brownian motion.
Hence its position at $t+\Delta t$ is normally distributed around  $X(t)$,
\begin{align} \label{eq:X_propagator_untethered}
    P\left(X(t+\Delta t)\right)
    =
    \frac{1}{4\pi D \Delta t}
    \exp \left( -\tfrac{\left(X(t+\Delta t)-X(t)\right)^2}{4D\Delta t} \right).
\end{align}
In the tethered state $S(t) = 1$, the probability density function of $X(t + \Delta t)$ is given by the solution of the Fokker-Planck equation with strong friction in a harmonic potential around the anchor point $X^*(t)$ \cite{Reichl2016},
\begin{align} \label{eq:X_propagator_tethered}
    &P\left(X(t+\Delta t)\right)
    = \\
    &\frac{1}{2\pi A'(\Delta t)} \exp\left({-\tfrac{\big(X(t + \Delta t)-\phi(\Delta t) X(t)-(1-\phi(\Delta t))X^{*}(t)\big)^2}{2A'(\Delta t)}}\right),  \nonumber
\end{align}
where we defined two auxiliary variables:
\begin{align}
    \phi(\Delta t) &= \exp{\left(-\frac{D\Delta t}{A}\right)}, &
    A'(\Delta t)&=(1-\phi^2(\Delta t))A.
    \label{eq:auxiliary variables}
\end{align}

\subsection{Discretized dynamics} \label{sec:discretized_dynamics}
The model defined above describes the full continuous-time dynamics.
In principle, we could use it to estimate the model parameters $\Theta = (\taustick,\tauunstick,D,A)$.
However, the experimental setup poses two difficulties: First, we can only measure the particle positions $X(t)$ and do not have access to the particle states $S(t)$ and tether points $X^*(t)$. This is the core challenge of the problem which we address in \cref{sec:problem algorithm}. Second, we only sample the process at discrete times $t_1, t_2, \dots, t_N$, separated by a finite time resolution $\Delta t=t_i-t_{i-1}$.
To account for this, we define a discrete Markov process, analogous to the continuous process, in which the tether point is constrained to be one of the previously observed positions.
This discrete Markov process defines a sequence of random states $\{F_n\}_{n=1}^N $ where
\begin{align}
    F_n := F(t_n) = \big(X(t_n),\ S(t_n),\ X^*(t_n)\big).
\end{align}
The transition probabilities of $F_n$ are given by the product of the transition probabilities of $S$ and $X$:
\begin{align}
    P(F_{n+1}|F_n)&=P(S_{n+1}|S_n;\Theta)P(X_{n+1}|F_n ;\Theta),
    \label{eq:discretized_step_dynamics}
\end{align}
where $X^*_{n+1}$ is uniquely determined by the history of $\{X_n\}$ and $\{S_n\}$, in accordance with \cref{eq:continuous xstar}:
\begin{align} \label{eq:X_nplus1_star}
    X^*_{n+1}=
    \begin{cases}
        X_{n+1} & S_{n+1}=1 \mbox{ and } S_{n} = 0, \\
        X_{n}^* & \mbox{otherwise}.
    \end{cases}
\end{align}
The state transitions are Poissonian and to leading order in $\Delta t$ read \begin{align}
    \label{eq: S_propagator}
    P&\left(S_{n+1}| S_n\right) =
    \begin{cases}
        1-\frac{\Delta t}{\tau_n} &   S_{n+1}=S_n \\
        \frac{\Delta t}{\tau_n} & S_{n+1}\ne S_n,
    \end{cases}
\end{align}
where $\tau_n=\tau_0$ if $S_n=0$ and $\tau_n=\tau_1$ if $S_n=1$.
The distribution of the particle position in the next step follows Eq. \eqref{eq:X_propagator_untethered} and \eqref{eq:X_propagator_tethered},
\begin{align}
    \label{eq: X propagator full}
    P(X_{n+1}&|F_n; \Theta)\\
    &=\begin{cases}
          \frac{1}{4\pi D \Delta t}e^{-\frac{1}{4D\Delta t}\Big(X_{n+1}-X_n\Big)^2} & S_n=0,\\
          \frac{1}{2\pi A'}e^{-\frac{1}{2A'}\Big(X_{n+1}-\phi X_n-(1-\phi)X^{*}_n\Big)^2} & S_n=1.
    \end{cases} \nonumber
\end{align}
where now $\phi$ and $A'$ are constants that depend on $A, D$ and the time step $\Delta t$ as in \cref{eq:auxiliary variables}.
Furthermore, we assume that at the beginning of the measurement the particle is either free ($S_1=0$) or tethered at $X_1$ ($S_1=1, X^*_1=X_1$). Importantly, $X^*_n$ can only attain values among the previously visited positions $X_1,\dots,X_n$. 

\begin{figure}
\includegraphics[width=1.002\linewidth]{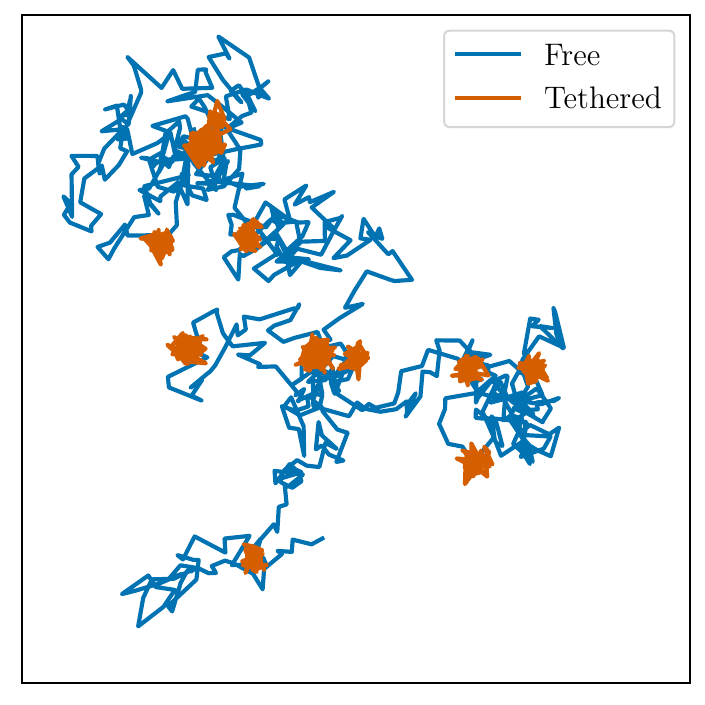}
\caption{An example trajectory of $N=2000$ steps generated from the discrete model with $\taustick=\tauunstick=100$, $D=1$ and $A=0.5$. The sampling time step is $\Delta t = 1$. The trajectory is color-coded according to the particle's state, where blue and orange depict the free and tethered states, respectively.}

\label{fig: example trajectory}
\end{figure}

As usual for Markov processes, the probability of a trajectory $\{F_n\}_{n=1}^N$ is the product of the transition probabilities in each step, thus the log-probability is additive:
\begin{align}
    \begin{split}
        \log&{P\left(\{F_n\}_{n=1}^N;\Theta\right)} \\
        &= \log P(F_1) + \sum_{n=1}^{N-1}{\log{P(F_{n+1}|F_{n}; \Theta)}} . 
    \end{split}
    \label{eq:full trajectory probability}
\end{align}
An example trajectory generated from the discrete model is depicted in \cref{fig: example trajectory}.

\subsection{Time scales}
\begin{figure}
    \includegraphics[width=\linewidth]{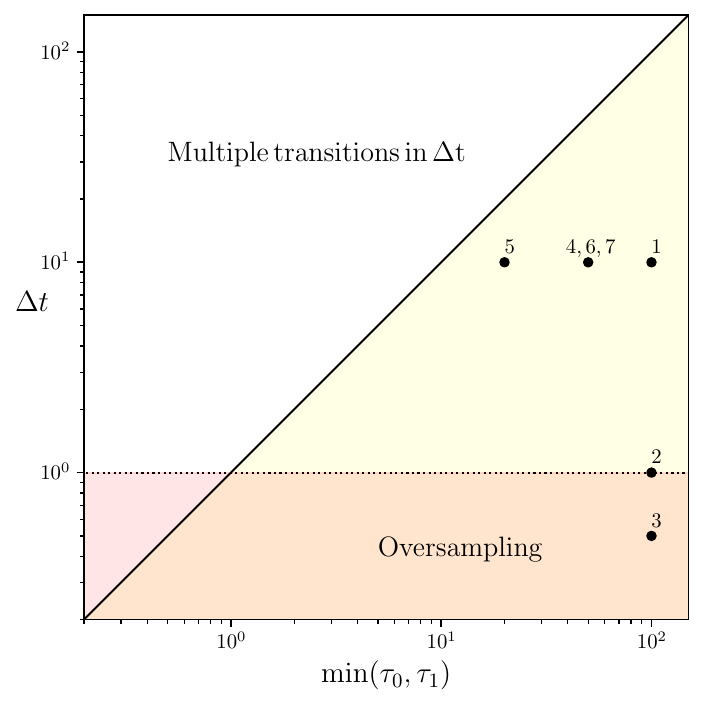}
    \caption{A section of the phase space of $\Delta t, \taustick, \tauunstick$, in time units of $\frac{A}{D}$. Note the logarithmic scale. The yellow region is where our assumptions hold. The bottom horizontal boundary of the yellow region corresponds to the inequality $\frac{A}{D}\lesssim \Delta t$ and the diagonal boundary corresponds to the inequality $\Delta t \ll \taustick,\tauunstick$. The bottom and left regions are annotated according to the discussion in this subsection. The markers correspond to the seven regimes to be analyzed in \cref{sec:results}.}
    \label{fig: triangle}
\end{figure}
As seen above, the problem involves multiple time scales. Three of them are given by the model parameters and represent the underlying physics: $\taustick,\tauunstick$, and $\frac{A}{D}$ which is the equilibration time of the Brownian particle with the harmonic potential. The other two time scales, $\Delta t$ and the total experiment duration $T=t_N-t_1$ are properties of the experiment.
We employ several realistic working assumptions about these time scales that greatly reduce computational complexity. First, we assume that $\taustick,\tauunstick \gg \frac{A}{D}$, which physically means that the rate of tethering/untethering events is much smaller than the inverse of the harmonic equilibration time of the particle. Violation of this condition means that the particle can untether before the tethering potential has a significant effect and thus tethering events will not be experimentally discernible. This regime corresponds to standard Brownian motion with an effective diffusion constant smaller than $D$.
Second, we require that $\Delta t\ll \taustick, \tauunstick$, to avoid the possibility of multiple transitions of tethering or untethering events within a single sampling interval. This assumption is experimentally realistic since modern cameras can easily achieve frame rates larger than $10^3$Hz, and in many experiments, $\taustick$ and $\tauunstick$ are of orders of at least seconds~\cite{Chakraborty2019,Xu2011,Callegari2019}. In any case, if $\Delta t\approx \taustick, \tauunstick$, this would again correspond to an essentially pure diffusive behavior in the discretized data.
Third, if $\Delta t<\frac{A}{D}$, the time discretization resolves the equilibration of a tethered particle with its confining potential. Since our goal is only to extract the physical parameters of the system, such resolution does not add relevant information and only increases the computational cost. Therefore, under-sampling the discrete dynamics to increase $\Delta t$ should not lead to a significant loss of accuracy in the estimation of $\taustick,\tauunstick$ but would greatly reduce the search space. This is explicitly demonstrated in~\cref{sec:results}. Even if the trajectories are experimentally measured with small $\Delta t$, we can safely downsample them such that $\frac{A}{D}\lesssim \Delta t$, or in other words $\phi\lesssim e^{-1}$, cf.~\cref{eq:auxiliary variables}.
For concreteness, we mention the experimental parameters of the diffusing nanoparticles system of Chakraborty et. al. \cite{Chakraborty2019}. In this system, $\taustick,\tauunstick \sim 1\text{s}$, $D\sim10\mu$m$^2$/s, $A\sim 1\mu\text{m}^2$, which means a sampling interval of $\Delta t = A/D = 0.1$s is sufficient and is easily achievable experimentally.
To conclude this discussion, we assume the following separation of time scales:
\begin{equation}
    \frac{A}{D} \lesssim \Delta t \ll \taustick,\tauunstick,
    \label{eq: time constraints}
\end{equation}
The leading-order expansion of \cref{eq: X propagator full}, which is first order in~$\Delta t$ and zeroth order in~$\phi$, now reads
\begin{align}
    \label{eq: propagators}
    P&\left(S_{n+1}| S_n\right) =
    \begin{cases}
        1-\frac{\Delta t}{\tau_n} &   S_{n+1}=S_n \\
        \frac{\Delta t}{\tau_n} & S_{n+1}\ne S_n
    \end{cases}\\
    \nonumber 
    P&\left(X_{n+1}|F_n\right) =
    \begin{cases}
          \frac{1}{4\pi D \Delta t}e^{-\frac{1}{4D\Delta t}\Big(X_{n+1}-X_n\Big)^2} & S_n=0\\
          \frac{1}{2\pi A}e^{-\frac{1}{2A}\Big(X_{n+1}-X^{*}_n\Big)^2} & S_n=1,
    \end{cases} 
\end{align}
where $\tau_n=\tau_0$ if $S_n=0$ and $\tau_n=\tau_1$ if $S_n=1$.

To conclude the discussion regarding time scales, \cref{fig: triangle} illustrates the section of phase space where our assumptions hold (in time units of $\frac{A}{D}$). For a point in the yellow region in the figure, the farther it is from the two boundary lines, the better is the separation of scales and the better our assumptions hold. The marked points correspond to regimes that will be analyzed in \cref{sec:results}.

\section{Our method}
\label{sec:problem algorithm}
Our problem is as follows: in an experiment, we can measure the observed states $\{X_n\}$, but we do not have access to the hidden states $\{S_n, X^*_n\}$, nor to the model parameters $\Theta=(\taustick,\tauunstick,D,A)$. We use the term {\it hidden path} to denote the sequence of the hidden states $\{S_n\}_{n=1}^N$ from which the sequence $\{X^*_n\}_{n=1}^N$ can be determined. The goal is to infer the model parameters $\Theta$ from a series of measurements. To this end, we developed an alternating maximization algorithm \cite{Li2019}, similar to the classical expectation-maximization (EM) algorithm \cite{Dempster1977}, to estimate both the hidden states and the parameters of our hidden Markov model.

If the hidden path is known, the problem of optimal parameter estimation is fairly standard and is typically solved by maximizing the likelihood of the model parameters, which according to Bayes' rule is proportional to the exponential of \cref{eq:full trajectory probability}. Maximizing $P\left(\Theta | \{F_n\}\right)$ can be done numerically or using analytical approximations, as described below. However, when the hidden path is not known, the likelihood function to consider is
\begin{align}
    \mathcal{L}\left(\Theta|\{X_n\}\right)&=\frac{P(\Theta)}{P(\{X_n\})}P\left(\{X_n\}\right|\Theta) && \text{(Bayes)}
    \nonumber\\  &=\frac{P(\Theta)}{P(\{X_n\})}\sum_{\{S_n\}}P\left(\{F_n\}\right|\Theta),
    \label{eq:likelihood_function}
\end{align}
where the sum is over all $2^N$ possibilities for the hidden paths, $P(\Theta)$ is the prior probability distribution for the model parameters, and the evidence $P(\{X_n\})$ is a constant we may ignore \cite{MacKay2006}. Recall that $\{F_n\}=\{X_n, S_n, X^*_n\}$ is the trajectory of both the observed and hidden states of the particle, and the log-probability of such a trajectory is the sum of the log-probabilities of all the steps, as in \cref{eq:full trajectory probability}.
Computing the sum in \cref{eq:likelihood_function} is intractable for typical values of~$N$.
However, numerical evidence shows that most hidden paths are very unlikely and thus have a negligible contribution to this sum. Taking a uniform prior, $P(\Theta) = {\rm const}$, we posit that
\begin{equation}
    \log{\mathcal{L}(\Theta|\{X_n\})} \sim \max_{\{S_n\}}{\log P(\{F_n\}|\Theta)}.
    \label{eq:saddlepointlikelihood}
\end{equation}
A key idea of our method is to use the RHS of Eq.~\eqref{eq:saddlepointlikelihood} as a proxy for the computationally intractable LHS.
It is analogous to the saddle-point approximation from statistical mechanics, where the integral is replaced with the maximum of the integrand.
To compute the RHS of Eq.~\eqref{eq:saddlepointlikelihood}, we need only to find the most likely hidden path $\{\hat{S}_n\}_{n=1}^N$, given the model parameters $\Theta$. This discrete optimization problem can be efficiently solved using the Viterbi algorithm from dynamic programming~\cite{Viterbi67,MacKay2006}.
Below we briefly describe the parameter estimation method and how the Viterbi algorithm is implemented for our model. Then we present our alternating maximization approach for estimating the maximum likelihood model parameters.

\subsection{Parameter estimation}\label{parameter_estimation}
Given the most likely hidden path $\{\hat{S}_n\}$, maximizing the log-likelihood in \cref{eq:saddlepointlikelihood} is maximizing a sum of the log terms from \cref{eq: propagators}. We use the following maximum-likelihood estimators (MLE) to approximate the most likely model parameters:
\begin{align}
    \hat{\tau}_{0} &:= \frac{N_{00}+N_{01}}{N_{01}} \Delta t \qquad
    \hat{\tau}_{1} := \frac{N_{11}+N_{10}}{N_{10}}\Delta t  \nonumber\\  
    \hat{D} &:= \frac{1}{4(N_{00}+N_{01})\Delta t}  \sum_{n=1}^{N-1}{(1-\hat{S}_{n})(X_{n+1}-X_{n})^2} \nonumber\\
    \hat{A} &:=\frac{1}{2(N_{10}+N_{11})}\sum_{n=1}^{N-1}{\hat{S}_{n}(X_{n+1}-X^{*}_{n})^2},
\label{eq: most likely parameters}
\end{align}
where $N_{ij}$ is the number of $i\to j$ transitions of $\hat{S}_n$ that occur along the trajectory. The derivation of these estimators is trivial by taking the derivative of \cref{eq:saddlepointlikelihood} with respect to each model parameter and equating to zero.

\subsection{Finding the most likely sequence of states}
\label{subsec: finding the hidden states}
\begin{figure}
\hspace*{-19pt}\includegraphics[width=1.12\linewidth]{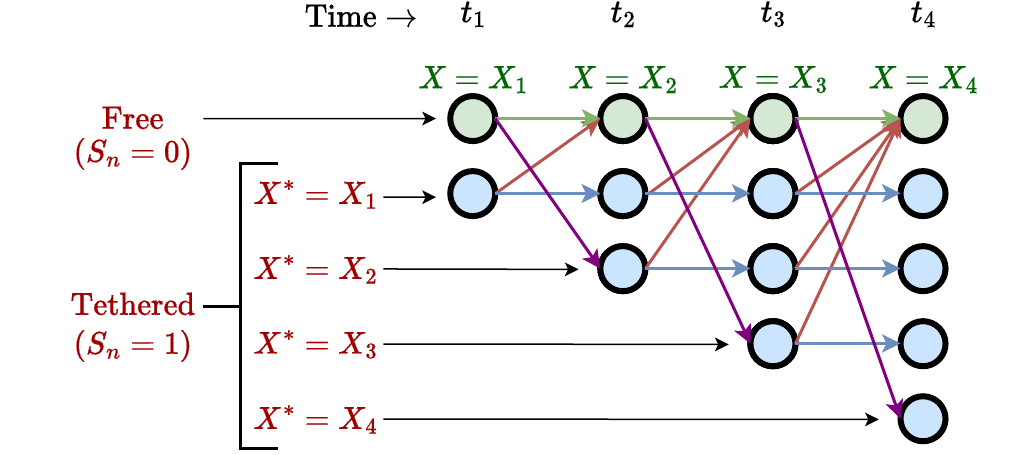}
\caption{A trellis graph of the observed and hidden states, given the observed particle positions $X_1,...,X_4$. Each node represents a different state $F_n = (X_n,S_n,X^*_n)$, with the row corresponding to the hidden state, and the column to the time $n$ and the observed state $X_n$. The edges represent allowed state transitions and are weighted according to the logarithm of \cref{eq:discretized_step_dynamics}. Each trajectory $\{F_n\}$ is given as a path on the graph that advances from left to right.
}
\label{fig: trellis}
\end{figure}

We now describe how to find the maximum likelihood path $\{\hat{S}_n\}$, conditioned on the model parameters $\Theta$.
Recall that the tether point $X^*_n$ can only assume one of the previously visited positions $X_1,X_2,...,X_n$, see~\cref{eq:X_nplus1_star}.
We represent the set of all possible paths as a directed layer graph (\cref{fig: trellis}), known as a \emph{trellis} in the Viterbi literature.
In this graph, columns correspond to the discrete time~$n$ and rows to the tether point $X^*_n$ which can be either one of $X_1, \ldots, X_n$ or \emph{free} (untethered).
Each vertex represents the state of the particle at time $n$ and each path from left to right corresponds to a specific sequence of tethered/free states.

We set the edge weights to be the logarithm of the transition probabilities between states according to~\cref{eq:discretized_step_dynamics}.
With this choice of edge weights, the log-likelihood of a path (\cref{eq:full trajectory probability}) is just $\log P(F_1)$ plus the sum of edge weights along the corresponding path in the graph.
Thus, finding the maximum-likelihood sequence of hidden states is reduced to the problem of finding the path of maximum weight in a directed layer graph.
The latter problem is efficiently solved using the Viterbi algorithm~\cite{Viterbi67}. This algorithm scans the trellis column by column from left to right and computes, for each vertex, the maximum-weight path that ends at that vertex.

\subsection{The alternating maximization algorithm} \label{em_algorithm}
Given an estimate of the hidden path $\{S_n\}$, we can apply \cref{eq: most likely parameters} to obtain the maximum likelihood estimate of the model parameters $\Theta=(\taustick,\tauunstick,D,A)$. Conversely, given an estimate of the model parameters $\Theta$, it is easy to find the maximum likelihood hidden path using the Viterbi algorithm as explained in the previous subsection.
Combining these two observations naturally leads to an alternating maximization procedure for estimating both the most likely model parameters $\hat{\Theta}$  and the most likely hidden path:
\begin{enumerate}
    \item {\bf Initial guess}: Guess an initial value $\hat{\Theta}^{(0)}$ for the four model parameters.

    \item {\bf Path maximization step}: Conditioned on  the current parameter estimate $\hat{\Theta}^{(m)}$, apply the Viterbi algorithm to find $\{S_n^{(m)}\}_{n=1}^N$, the most likely hidden path given $\hat{\Theta}^{(m)}$,
    \begin{align}
        S^{(m)} := \argmax_{\{S_n\}} \mathcal L \big(\{S_n\} | \hat{\Theta}^{(m)}, \{X_n\}\big).
    \end{align}

    \item {\bf Parameter maximization step}: Use \cref{eq: most likely parameters} to obtain the maximum likelihood estimate of the model parameters conditioned on the current estimate of the hidden path,
    \begin{align}
        \hat{\Theta}^{(m+1)}
        :=
        \hat{\Theta} \big( \{S^{(m)}\}, \{X_n\} \big).
    \end{align}

    \item {\bf Convergence}: Alternate between steps 2 and 3 until $\left|\hat{\theta}^{(m+1)}-\hat{\theta}^{(m)}\right|/\hat{\theta}^{(m)} \le \epsilon$ for each $\hat\theta\in\hat\Theta$.
 \end{enumerate}

Since the sample space is discrete, convergence typically occurs exactly, i.e.~$\hat{\Theta}^{(n+1)}=\hat{\Theta}^{(n)}$. However, in our experiments we used $\epsilon = 10^{-3}$ to stop the iterations when the relative change is small. The specific threshold $10^{-3}$ is inconsequential.
Furthermore, we set the maximum number of iterations to 20, to prevent the possibility of infinite loops that alternate between several discrete hidden paths without satisfying the convergence criterion (in practice, less than 10 iterations typically suffice for convergence).
We also define a criterion for divergence: if after the parameter maximization step the estimators $\hat{\tau}_0$ or $\hat{\tau}_1$ exceed the arbitrary threshold of $0.9T$, we say the algorithm diverged and stop the iterations. This is done to avoid cases where the particle is estimated to stay tethered or untethered for the entire sampling time due to an unphysical divergence of the MLE estimators. In practice, we test the regimes where $T\gg \taustick,\tauunstick$ and most runs do not diverge. We say the algorithm has converged if the convergence criterion was satisfied without first triggering the divergence criterion.

The Viterbi algorithm scans each edge on the graph once, leading to an $O(N^2)$ complexity. However most paths are not likely, e.g. a path where at some point $S_n=1$ and $(X_n-X^*_n)^2\gg A$, meaning the particle roams far from its tether point.
To improve the running time, we implemented edge pruning of the trellis graph.
Specifically, we discard the paths of all but the $q$ most likely tethered nodes at each time step (column) during the algorithm's execution. This reduces the complexity to $O(qN)$. In our tests, we used $q=10$, which showed no significant impact on the results compared to no pruning.

In \cref{sec:k-most} we discuss an extension of our algorithm that outputs the $K$ most likely hidden paths rather than the single best path. We generally found that the $100$ most likely paths yield similar parameter estimates, so the results in the rest of this paper use $K=1$.

\subsection{\label{sec:bootstrap} Bootstrap bias correction}
For reasons that will be discussed in \cref{sec:results}, the MLE predictions for the model parameters give very good estimates of $D$ and $A$, but consistently overestimate the waiting times $\taustick$ and $\tauunstick$. This overestimation can be largely corrected by using a parametric bootstrap procedure \cite{Efron1994}. The idea is simple: define the bias $\Delta$ as the expected difference between the MLE and the true parameters: $\Delta=\left\langle\hat \Theta\right\rangle -\Theta$. Assuming the bias does not depend sensitively on $\Theta$ itself, one can estimate $\Delta$ by simulating trajectories with parameters $\hat \Theta$ and rerunning the procedure on this simulated data (for which we know the true parameters).
This is done as follows:
\begin{enumerate}
    \item Apply the alternating maximization algorithm described in Section \ref{em_algorithm}  to obtain the MLE $\hat \Theta$.
    \item Generate $M$ random trajectories using $\hat \Theta$ as the ``true'' simulated model parameters.
    \item Apply the alternating maximization algorithm to each simulated trajectory, yielding the estimated model parameters $\hat \Theta'_i$ for $1\le i \le M$.
    \item The bias is estimated by the median value of the observed biases,
    $$\hat{\Delta} := \operatornamewithlimits{median}_{1\le i \le M}(\hat \Theta'_i-\hat \Theta)\ .$$
    \item Finally, the bias is used to correct the original estimate. That is, the bias-corrected estimate for the true model parameters is given by
    \begin{align}
        \label{eq:bootstrap}
        \hat \Theta^B = \hat\Theta-\hat{\Delta}.
    \end{align}
\end{enumerate} 
Note that the median is calculated for each parameter $\theta\in\Theta$ independently. We chose to estimate $\hat{\Delta}$ using the median since it is robust to outliers and we found that it gave better results in mean squared error.

\section{\label{sec:results} Results}
We test the algorithm's performance on synthetically generated trajectories, whose model parameters $\Theta$ are known. The implementation of the algorithm in Python, as well as the code to generate all figures in this manuscript, is available at~\cite{github}.

In all trajectories, we use $D=1$ and $A=1$ and a total experiment duration of $T=10000$. Note that both $D$ and $A$ can always be set to unity by properly choosing length and time units, so this choice is without loss of generality. Thus, each experiment is specified by three scalars: the sampling time $\Delta t$ and the model parameters $\taustick$,$\tauunstick$, all expressed in time units of $\frac{A}{D}=1$. We focus on seven different regimes of $\Theta$ and $\Delta t$, detailed in \cref{table:regimes}. We mention that the conditions in \cref{eq: time constraints} are barely satisfied in regimes $3$ and $5$. This is intentional as we wish to test the algorithm slightly beyond its bounds.

\newcolumntype{x}[1]{>{\raggedleft\arraybackslash\hspace{0pt}}p{#1}}
\begin{table*}
    \renewcommand{\arraystretch}{1.3}
    \begin{center}
        \begin{tabular}{lx{1cm}x{1cm}x{1cm}x{1cm}x{2cm}x{2cm}x{2cm}x{2cm}x{2cm}}
            {Regime} & {$D,A$} & {$\Delta t$} & {$\taustick$} & {$\tauunstick$} & {Acc. (\%)} & {$\hat{\tau}_0$} & {$\hat{\tau}_1$} & $\hat{D}$ & $\hat{A}$  \\\toprule 
            \textbf{1} & $1$ & $10$ & $100$ & $100$ & $96\pm 2$ & $131 \pm 24$ & $130\pm 19$ & $1.00 \pm 0.05$ & $0.99 \pm 0.05$
            \\\hline
            \textbf{2} & $1$ & $1$ & $100$ & $100$ & $94\pm 2$ & $122 \pm 21$ & $122\pm 16$ & $1.00 \pm 0.01$ & $0.99 \pm 0.02$
            \\\hline
            \textbf{3} & $1$ & $0.5$ & $100$ & $100$ & $88\pm 4$ & $125 \pm 22$ & $123\pm 17$ & $1.00 \pm 0.01$ & $0.99 \pm 0.02$  
            \\\hline
            \textbf{4} & $1$ & $10$ & $50$ & $50$ & $93\pm 2$ & $77 \pm 11$ & $75\pm 8$ & $0.99 \pm 0.05$ & $0.98 \pm 0.05$  
            \\\hline
            \textbf{5} & $1$ & $10$ & $20$ & $20$ & $87\pm 2$ & $47 \pm 9$ & $43\pm 5$ & $0.97 \pm 0.06$ & $0.94 \pm 0.06$  
            \\\hline
            \textbf{6} & $1$ & $10$ & $200$ & $50$ & $96\pm 1$ & $356 \pm 95$ & $79\pm 12$ & $0.99 \pm 0.04$ & $0.97 \pm 0.09$  
            \\\hline
            \textbf{7} & $1$ & $10$ & $50$ & $200$ & $97\pm 2$ & $60 \pm 11$ & $248\pm 43$ & $1.01 \pm 0.08$ & $1.00 \pm 0.04$
        \end{tabular}
    \end{center}
    \caption{True model parameters, accuracy, and estimated parameters (mean $\pm$ standard deviation) for the 7 regimes analyzed using the alternating maximization algorithm (Section~\ref{em_algorithm}) without bias-correction. The accuracy is defined as the fraction of time steps at which the algorithm correctly predicted both the tethered state $S$ and the tether point $X^*$. 
    }
    \label{table:regimes}
\end{table*}

\subsection{Stability}
\begin{figure*}
\includegraphics[width=\linewidth]{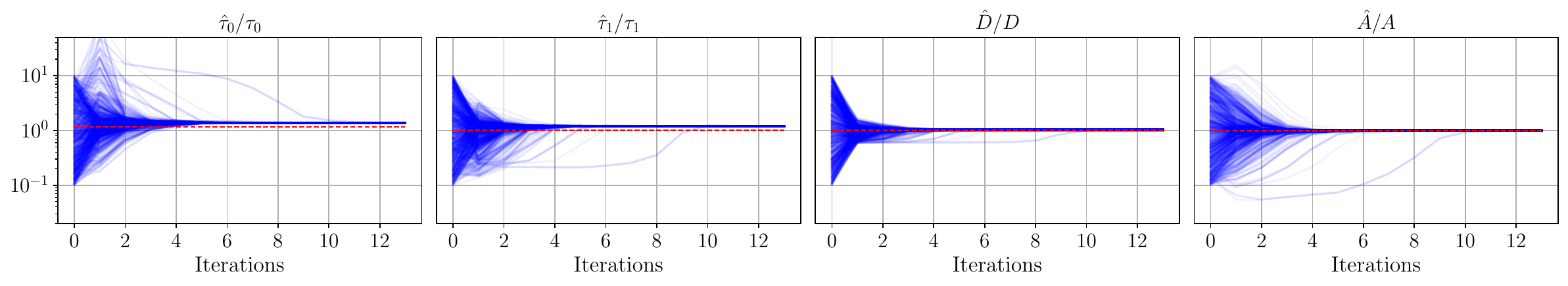}
\caption{\textbf{The algorithm is insensitive to the initialization $\Theta^{(0)}$:} Here we show the evolution of the estimators $\hat{\tau}_0,\hat{\tau}_1,\hat{D},\hat{A}$ of the four model parameters along the algorithm iterations over a single synthetic trajectory. The values are normalized by the true parameter values so that a perfect prediction would be unity in all panels. The first iteration is the initial parameter guess. 
We drew 1000 initial guesses at random from a log-uniform distribution around the true value and ran the algorithm, keeping only the 960 runs that converged. The evolution of the estimators of each (converged) run is shown, exhibiting a strong convergence towards a single fixed point. Note that the vertical axis is logarithmic. The dashed red line represents the most likely parameter value, given the true hidden path. It does not coincide, in general, with the true model parameters.}
\label{fig: indifference_to_init_conditions}
\end{figure*}

As described above, the algorithm requires an initial guess of the model parameters $\hat{\Theta} ^{(0)}$. We found empirically that the algorithm converges to a single fixed point almost regardless of the initial parameters. To demonstrate this, we show in \cref{fig: indifference_to_init_conditions} the results of the algorithm when applied to a single observed trajectory corresponding to regime $1$, with 1000 different $\hat{\Theta} ^{(0)}$'s, randomly drawn from a log-uniform distribution spanning two decades around $\Theta$.
We mention this experiment's wide distribution of initializations $\hat{\Theta} ^{(0)}$ is intended to illustrate the algorithm's stability, and in a practical situation, better priors can often be used, especially for $D$ and $A$.
The convergence for other regimes is qualitatively similar and is not shown here. Out of the 1000 algorithm runs, 960 converged and only these runs were taken into account in the figure. 
Since the outcome of the algorithm is largely independent of the initial guess, in what follows we use the true model parameters $\Theta$ as the initial guess $\hat{\Theta}^{(0)}$.

\subsection{Recovering the parameters and hidden states}
For each of the regimes in \cref{table:regimes}, we generated 1000 synthetic trajectories and ran the algorithm with initial parameters that are equal to the true model parameters, as discussed in the preceding subsection regarding stability. Over $98\%$ of the runs in each regime have converged (the other runs entered a loop and stopped after the maximum number of iterations). Runs that converged typically did so within 3-8 iterations.
We define the accuracy as the fraction of time steps for which the algorithm predicted both the correct state $S_n$ and tether point $X^*_n$ (if $S_n=1$). The mean and standard deviation of the accuracy over all converged runs are detailed in \cref{table:regimes}. It is seen that the accuracy is fairly high in all regimes, and that it decreases as $\Delta t$ approaches $\frac{A}{D}$ from above or $\taustick$,$\tauunstick$ from below, consistently with the constraints of \cref{eq: time constraints}.

Next, we examine how well the algorithm recovers the model parameters. For each regime, \cref{table:regimes} depicts the mean and the standard deviation of the algorithm's estimates $\hat{\Theta}=(\hat{\tau}_0,\hat{\tau}_1,\hat{D},\hat{A})$ over all converged runs. These results exhibit two clear trends: the temporal parameters $\taustick,\tauunstick$ are consistently overestimated, while the spatial ones $D,A$ are correctly estimated (although there is a very slight yet consistent underestimate of $A$ in most regimes). The overestimate of the temporal parameters ranges from $20\%$ to $120\%$ and will be discussed in \cref{sec:overestimation}.

\begin{figure*}
    \includegraphics[width=\linewidth]{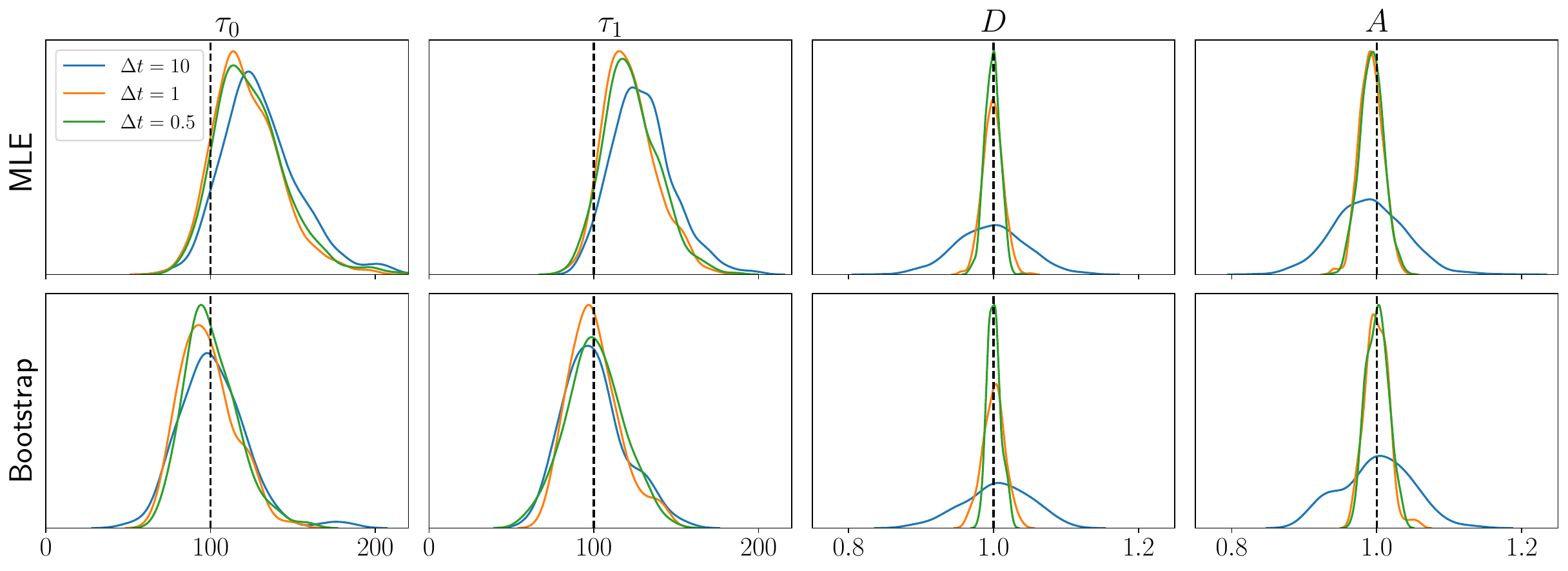}
    \caption{Distribution of the algorithm's estimates of physical model parameters for regimes 1,2,3 with $\taustick=\tauunstick=100$ and $\Delta t = 10,1,0.5$, respectively. The dashed lines represent the true parameter value. Top row: the maximum likelihood estimates of the alternating maximization algorithm. Bottom row: the estimators after the bootstrap bias correction procedure.}
    \label{fig: params_dt_dependence}
\end{figure*}

To investigate the effect of the sampling time $\Delta t$, we focus on the three regimes $1,2,3$ which differ only by $\Delta t$ (see  \cref{table:regimes}). The distributions of the estimated model parameters are depicted in the top row of \cref{fig: params_dt_dependence}. 
Note that the centers of the distributions hardly change with~$\Delta t$.
This supports the claim described in \cref{sec:model} that the sampling time can be increased without losing information, as long as \cref{eq: time constraints} holds.
We mention that the width of the distribution for the spatial parameters $D,A$ decreases with $\Delta t$. That is to be expected since the accuracy of the MLEs increases with $N$, but for a fixed $T$, the number of samples $N$ is inversely proportional to~$\Delta t$.

By focusing on regimes $1,4,$ and $5$, the effect of $\taustick,\tauunstick$, when they are equal, can be isolated.
As $\taustick$ and $\tauunstick$ decrease, the accuracy decreases and the relative overestimate of $\taustick,\tauunstick$ increases. This is consistent with \cref{eq: time constraints}, since decreasing $\taustick,\tauunstick$ challenges the assumption that $\Delta t \ll \taustick,\tauunstick$.
Finally, regimes $6$ and $7$ explore the case of $\taustick\ne\tauunstick$. Note that when $\taustick<\tauunstick$, i.e.~when the particle is more likely to be tethered than free, the parameter estimates are better, with less overestimation for the temporal parameters, compared to the opposite case.

\subsection{Overestimation and bias correction}
\label{sec:overestimation}
As seen in \cref{table:regimes,fig: params_dt_dependence}, while the spatial parameters $D$ and $A$ are correctly estimated to high accuracy, the temporal parameters $\taustick,\tauunstick$ are consistently overestimated, with an increasing overestimation as $\taustick, \tauunstick$ decrease compared to $\Delta t$. 
This bias is due to a systematic misidentification of brief tethered or untethered intervals. That is, if the particle tethers and then untethers (or vice versa) over a short time window, the likelihood is dominated by the temporal terms (regarding $S_n$) in \cref{eq: propagators} over the spatial terms. As a result, the likelihood of a trajectory that stays in the same state is higher than that of a trajectory that switches tethering states twice within this short interval, even if the true dynamics involved such a switch.

To see this, take a short time interval of $l$ steps, in which the particle does not move much. Compare two sequences of hidden states, one in which the particle is tethered in the first step and untethered in the last one, and one in which it is free throughout the time window (a similar argument holds for the opposite scenario). Taking the difference between the log-likelihoods of the two hidden trajectories using  \cref{eq: propagators}, and expanding to leading order in $l$, we obtain
\begin{align}
    \Delta \mathcal{L} = 2\log\left(\frac{1-\Delta t/\tau}{\Delta t/\tau}\right) +\mathcal{O}\left(l, \Delta t^2\right)
\end{align}
where for simplicity we took $\taustick=\tauunstick=\tau$. For short time windows, this tends to be positive since $\Delta t \ll \taustick,\tauunstick$, with the first term slowly diverging as $\Delta t \to 0$.

In addition, since the distribution of the free and tethered interval durations times is exponential, it is dominated by such short intervals, which are systematically misclassified by the MLE estimates. An example of this phenomenon is illustrated in \cref{fig: K most likely}. Since the MLE estimates for $\taustick, \tauunstick$ (\cref{eq: most likely parameters}) are inversely proportional to the number of switching events, missing brief tethering/untethering intervals leads to over-estimation of the waiting times.
We correct this bias using a parametric bootstrap procedure, as described in Section \ref{sec:bootstrap}. The bottom row of \cref{fig: params_dt_dependence} shows the distribution of the bias-corrected estimates $\hat{\Theta}^B$, that are concentrated within a small region around the true value. Note that although the mode of the distributions for $\taustick, \tauunstick$ is slightly biased, the means of the distributions are close to the true values, as evident in \cref{table:bootstrap}. Importantly, the bootstrapping successfully mitigates the bias in the rate estimation without affecting the already accurate estimates of $D$ and $A$. In our simulations we generated $M=100$ bootstrapping trajectories (cf.~\cref{sec:bootstrap}) for each analyzed trajectory.

\begin{table*}
    \renewcommand{\arraystretch}{1.3}
    \begin{center}
        \begin{tabular}{lx{0.9cm}x{0.9cm}x{0.9cm}x{0.9cm}x{2.5cm}x{2.5cm}x{3cm}x{3cm}}
            {Regime} & {$D,A$} & {$\Delta t$} & {$\taustick$} & {$\tauunstick$} & {$\hat{\tau}_0^B$} & {$\hat{\tau}_1^B$} & $\hat{D}^B$ & $\hat{A}^B$  \\\toprule 
            \textbf{1} & $1$ & $10$ & $100$ & $100$ & $102\,(71-141)$ & $100\,(73-139)$ & $1.00\,(0.91-1.08)$ & $1.00\,(0.91-1.08)$
            \\\hline
            \textbf{2} & $1$ & $1$ & $100$ & $100$ & $98\,(73-129)$ & $100\,(76-138)$ & $1.00\,(0.97-1.03)$ & $1.00\,(0.97-1.03)$
            \\\hline
            \textbf{3} & $1$ & $0.5$ & $100$ & $100$ & $101\,(75-137)$ & $101\,(69-133)$ & $1.00\,(0.99-1.02)$ & $1.00\,(0.98-1.03)$  
            \\\hline
            \textbf{4} & $1$ & $10$ & $50$ & $50$ & $49\,(38-64)$ & $49\,(36-61)$ & $1.00\,(0.92-1.08)$ & $0.99\,(0.87-1.07)$  
            \\\hline
            \textbf{5} & $1$ & $10$ & $20$ & $20$ & $18\,(9-28)$ & $19\,(11-26)$ & $0.98\,(0.88-1.11)$ & $0.98\,(0.87-1.10)$  
            \\\hline
            \textbf{6} & $1$ & $10$ & $200$ & $50$ & $190\,(100-316)$ & $51\,(32-77)$ & $1.00\,(0.92-1.06)$ & $0.99\,(0.85-1.16)$  
            \\\hline
            \textbf{7} & $1$ & $10$ & $50$ & $200$ & $51\,(36-76)$ & $198\,(131-295)$ & $1.00\,(0.87-1.20)$ & $1.00\,(0.94-1.08)$
        \end{tabular}
    \end{center}
    \caption{True model parameters vs. their bootstrap bias-corrected estimates for the 7 regimes analyzed. The estimates display the mean and a $95\%$ confidence interval. We used $M=100$ synthetic bootstrap trajectories per analyzed trajectory.} 
    \label{table:bootstrap}
\end{table*}
\section{\label{sec:discussion} Discussion and outlook}
We developed and tested an efficient and accurate algorithm to analyze trajectories of diffusing particles with transient tethering, to identify tethering/untethering events, and to estimate the physical parameters of the system. The crux of our method is using the Viterbi algorithm to find the most likely sequence of hidden states, estimate the most likely model parameters for that sequence, and then bias-correct the estimates using a parametric bootstrap procedure. Our algorithm successfully recovers the model parameters and is largely insensitive to the parameter initialization.
It is applicable when the time interval between frames $\Delta t$ is significantly shorter than the typical tethering/untethering times and longer than, or comparable to, the equilibration time of the particle with the tethering potential. When the latter condition is not met, downsampling of the time series can be performed without significant loss of accuracy in estimating the physical parameters.

Our method as presented above only applies, of course, to normal 2D diffusion with Poissonian transient tethering. However, it can be readily generalized to other situations. For example, it can easily be generalized to a different spatial dimension ($d\ne2$) by adjusting the prefactors in \cref{eq: X propagator full}. In higher dimensions, the aforementioned misidentification of brief tethered intervals should be less prominent since a sequence of short steps becomes less likely (no recurrence). Other diffusive statistics (anomalous diffusion) or non-Poissonian tethering/untethering can also be modelled by adjusting the transition probabilities, \cref{eq: propagators}. For example, it could incorporate a drift term or an observational error term for position samples, as in Bernstein and Fricks's work \cite{Bernstein2016}. However, one must assume a specific functional form for the diffusion and tethering dynamics -- in the regime in which we are working, only a few dozen tethering/untethering events per trajectory are observed. Hence, due to the small sample size, statistical goodness-of-fit tests would lack sufficient power to distinguish, e.g., between Poissonian and non-Poissonian tethering. 

\begin{acknowledgments}
We thank Roy Beck, Yael Roichman, Indrani Chakraborty and Amandeep Sekhon for useful discussions and for providing experimental data.
YBS is supported by ISF grant 1907/22 and by Google Gift grant.
AM is supported by ISF grant 1662/22 and NSF-BSF grant 2022778.
\end{acknowledgments}

\appendix
\section{\label{sec:k-most} K most likely paths}
Our saddle-point approximation in \cref{eq:saddlepointlikelihood} replaces the sum over all hidden paths with the likelihood value of the single most likely path. An immediate generalization is to sum over the top $K$ likeliest paths, which would improve the estimation accuracy of $\mathcal{L}$.
Given the top $K$ likeliest paths, we can estimate the model parameters as a weighted sum of the MLEs of each path, cf.~\cref{eq: most likely parameters} as
\begin{equation}
    \hat{\Theta}^{[K]}=\frac{\sum_{i=1}^{K}{P(\{F_n\}^{[i]}|\Theta) \hat{\Theta}(\{F_n\}^{[i]})}}{\sum_{i=1}^{K}{P(\{F_n\}^{[i]}|\Theta)}} ,
    \label{K_most_likely_M_step}
\end{equation}
where $\{F_n\}^{[i]}$ is the trajectory corresponding to the $i$th most likely hidden path.
Finding the $K$ highest likelihood paths can be easily done by a slight modification of the Viterbi algorithm, also known as list Viterbi \cite{Seshadri1994, Roder2006}. The modification is discarding all but the $K$ most likely paths ending at each node, rather than all but the most likely path. Doing this increases the computational complexity by $K^2$.

In our numerical tests, we found no significant improvement in the results when increasing $K$ up to $K=100$. This is because the $K$ most likely paths are very similar, differing from each other in just a few steps, leading to very similar MLEs of $\Theta$. 
Typically, the top paths only differ by slight perturbations of the tethering and untethering times and do not display qualitative differences.

\begin{figure}

\includegraphics[width=\linewidth]{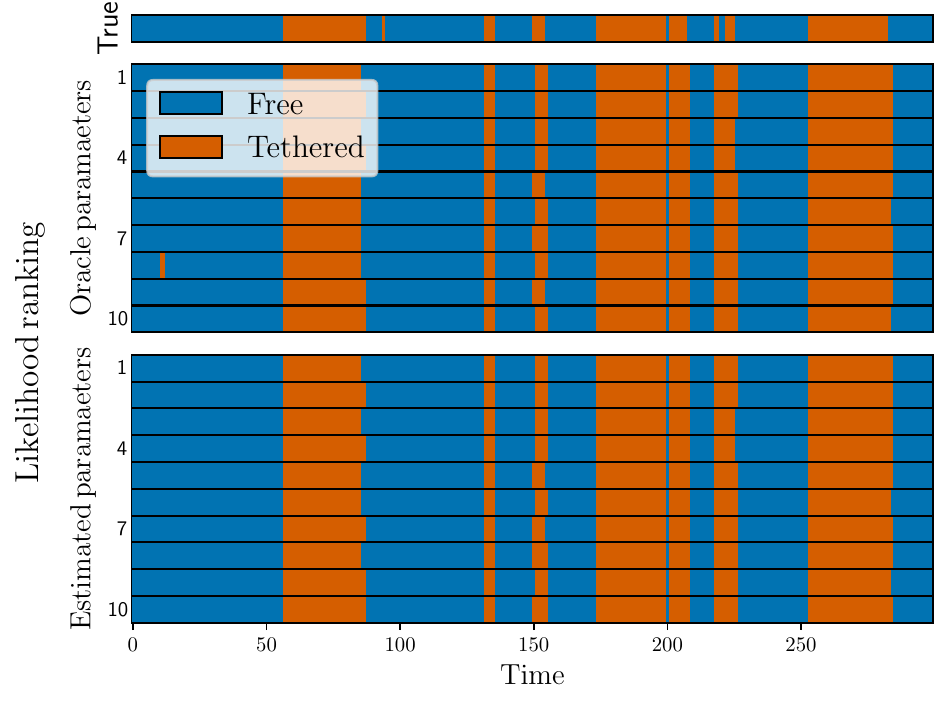}
\caption{The hidden paths of a single trajectory realization with $N=300$ steps, $\taustick=\tauunstick=100$, $D=1$, $A=1$, $\Delta t = 10$. The true hidden path is at the top. In the middle are the 10 most likely hidden paths conditioned on the true parameters, with the upper row being the most likely. At the bottom are the 10 most likely hidden paths given the algorithm's estimated model parameters, again with the upper row being the most likely. Blue intervals correspond to the free state, and orange intervals correspond to the tethered state.}
\label{fig: K most likely}
\end{figure}

To demonstrate this, \cref{fig: K most likely} depicts the true hidden path of a single trajectory along with the top $K=10$ most likely hidden paths. The likelihood of these paths can be computed either with respect to the most likely parameters given the true hidden path (``oracle''), or the estimated $\hat{\Theta}$ (without bootstrapping). It is seen that this choice does not significantly change the result, and  the most likely hidden paths are all qualitatively very similar, differing by just a few steps from each other, and are similar to the true hidden path. They miss the brief tethered interval at time $t=93$ and the brief free interval at time $t=220$.
Because the results were similar for $K$ up to 100, in the results section we presented only results pertaining to $K=1$. However, using $K>1$ top paths may be beneficial in other regimes.

\bibliography{references}
\end{document}